\documentclass[twocolumn]{aastex62}

\usepackage{graphicx}
\usepackage{float}

\newcommand{\Msun}{$M_{\odot}$}
\newcommand{\Mearth}{$M_{\oplus}$}

\newcommand{\Mdot}{$\dot{M}$}
\newcommand{\Mdust}{${M_{\rm{dust}}}$}
\newcommand{\Mdisk}{${M_{\rm{disk}}}$}

\newcommand{\msunyr}{$\rm{M_{\sun} \, yr^{-1}}$}

\newcommand{\mic}{$\mu$m}

\newcommand{\mstar}{$M_{*}$}

\newcommand{\brgamma}{Br$\gamma$}
\newcommand{\Lacc}{$L_{\rm{acc}}$}

\newcommand{\Lbrg}{$L_{\rm{Br\gamma}}$}

\newcommand{\mdotmdisk}{$\dot{M}$--$M_{\rm{disk}}$}
\newcommand{\tdisk}{$t_{\rm{disk}}$}

\begin{document}

\title{The $\dot{M}$--$M_{\rm{disk}}$ relationship for Herbig Ae/Be stars: a lifetime problem for disks with low masses?}

\correspondingauthor{Sierra L. Grant}
\email{sierrag@mpe.mpg.de}

\author[0000-0002-4022-4899]{Sierra L. Grant}
\affil{Max-Planck Institut f\"{u}r Extraterrestrische Physik (MPE), Giessenbachstr. 1, 85748, Garching, Germany}

\author[0000-0001-9524-3408]{Lucas M. Stapper}
\affil{Leiden Observatory, Leiden University, 2300 RA Leiden, the Netherlands}

\author[0000-0001-5217-537X]{Michiel R. Hogerheijde}
\affil{Leiden Observatory, Leiden University, 2300 RA Leiden, the Netherlands}
\affil{Anton Pannekoek Institute for Astronomy, University of Amsterdam, the Netherlands}

\author[0000-0001-7591-1907]{Ewine F. van Dishoeck}
\affil{Leiden Observatory, Leiden University, 2300 RA Leiden, the Netherlands}
\affil{Max-Planck Institut f\"{u}r Extraterrestrische Physik (MPE), Giessenbachstr. 1, 85748, Garching, Germany}

\author[0000-0001-5638-1330]{Sean Brittain}
\affil{Clemson University, 118 Kinard Laboratory, Clemson, SC 29631, USA}

\author[0000-0002-4147-3846]{Miguel Vioque}
\affil{Joint ALMA Observatory, Alonso de C\'ordova 3107, Vitacura, Santiago 763-0355, Chile}
\affil{National Radio Astronomy Observatory, 520 Edgemont Road, Charlottesville, VA 22903, USA}

\begin{abstract}
The accretion of material from protoplanetary disks onto their central stars is a fundamental process in the evolution of these systems and a key diagnostic in constraining the disk lifetime. We analyze the relationship between the stellar accretion rate and the disk mass in 32 intermediate-mass Herbig Ae/Be systems and compare them to their lower-mass counterparts, T Tauri stars. We find that the $\dot{M}$--$M_{\rm{disk}}$ relationship for Herbig Ae/Be stars is largely flat at $\sim$10$^{-7}$ M$_{\odot}$ yr$^{-1}$ across over three orders of magnitude in dust mass. While most of the sample follows the T Tauri trend, a subset of objects with high accretion rates and low dust masses are identified. These outliers (12 out of 32 sources) have an inferred disk lifetime of less than 0.01 Myr and are dominated by objects with low infrared excess. This outlier sample is likely identified in part by the bias in classifying Herbig Ae/Be stars, which requires evidence of accretion that can only be reliably measured above a rate of $\sim$10$^{-9}$ M$_{\odot}$ yr$^{-1}$ for these spectral types. If the disk masses are not underestimated and the accretion rates are not overestimated, this implies that these disks may be on the verge of dispersal, which may be due to efficient radial drift of material or outer disk depletion by photoevaporation and/or truncation by companions. This outlier sample likely represents a small subset of the larger young, intermediate-mass stellar population, the majority of which would have already stopped accreting and cleared their disks.  
\end{abstract}


\section{Introduction}\label{sec: intro}
Circumstellar disks are the birthplaces of planets and those planets must form in the first several million years of the disk lifetime before the disk dissipates. It is then important to understand how disks evolve and to characterize how that evolution impacts planet formation and vice versa. The rate at which material is being accreted onto the star from the disk and the disk mass are two key parameters in assessing the evolutionary state of a system. These two diagnostics probe different regions in the disk: the accretion rate traces the innermost star-disk connection and the disk mass traces the mass reservoir at tens to hundreds of au.

Despite the contrasting scales that the accretion rate (\Mdot) and disk mass (\Mdisk) probe, it has been predicted that the two quantities should be related and can give an estimate of the disk lifetime, $t_{\rm{disk}} = M_{\rm{disk}}/\dot{M}$ (e.g. \citealt{hartmann98, jones12, lodato17, rosotti17, sellek20,manara22}). The transfer of material inward from the outer disk can be affected by a variety of factors, including the formation of pressure traps, stellar irradiation and photoevaporation, MHD disk winds, and the presence of giant planets and companions (e.g., \citealt{jones12,rosotti17,tabone22,zagaria22}). Deviations from the nominal \mdotmdisk\ relationship can then indicate the presence of one or more of these processes. 

Recent observational efforts conducted at optical and near-infrared wavelengths paired with the numerous outer disk surveys, particularly with ALMA, have led to large populations of disks with both \Mdot\ and \Mdust\ measurements \citep{mendigutia12, manara16, manara20, ansdell17, mulders17,grant21,testi22,fiorellino22}. However, these surveys have greatly favored low-mass T Tauri stars, with the exception of \cite{mendigutia12} which was carried out before ALMA was operational. The more massive Herbig Ae/Be stars, by comparison, lack homogeneous (sub-)millimeter observations \citep{stapper22}, while they are well-covered in surveys focusing on accretion signatures (e.g., \citealt{donehew-brittain11, fairlamb15,fairlamb17,grant22,vioque22}). The disks around these intermediate-mass stars are thought to form giant exoplanets more efficiently than low-mass stars \citep{johnson10,reffert15}. Indeed, \cite{vandermarel21} use disk properties to tentatively point to a connection between stellar mass and giant planet formation. Therefore, it is essential to understand disk evolution and planet formation in the disks around intermediate-mass stars. In this work, we take these two key disk diagnostics, \Mdot\ and \Mdisk, to study the \mdotmdisk\ relationship in a sample of 32 Herbig Ae/Be objects.

\begin{figure*}
    \centering
    \includegraphics[scale=0.58]{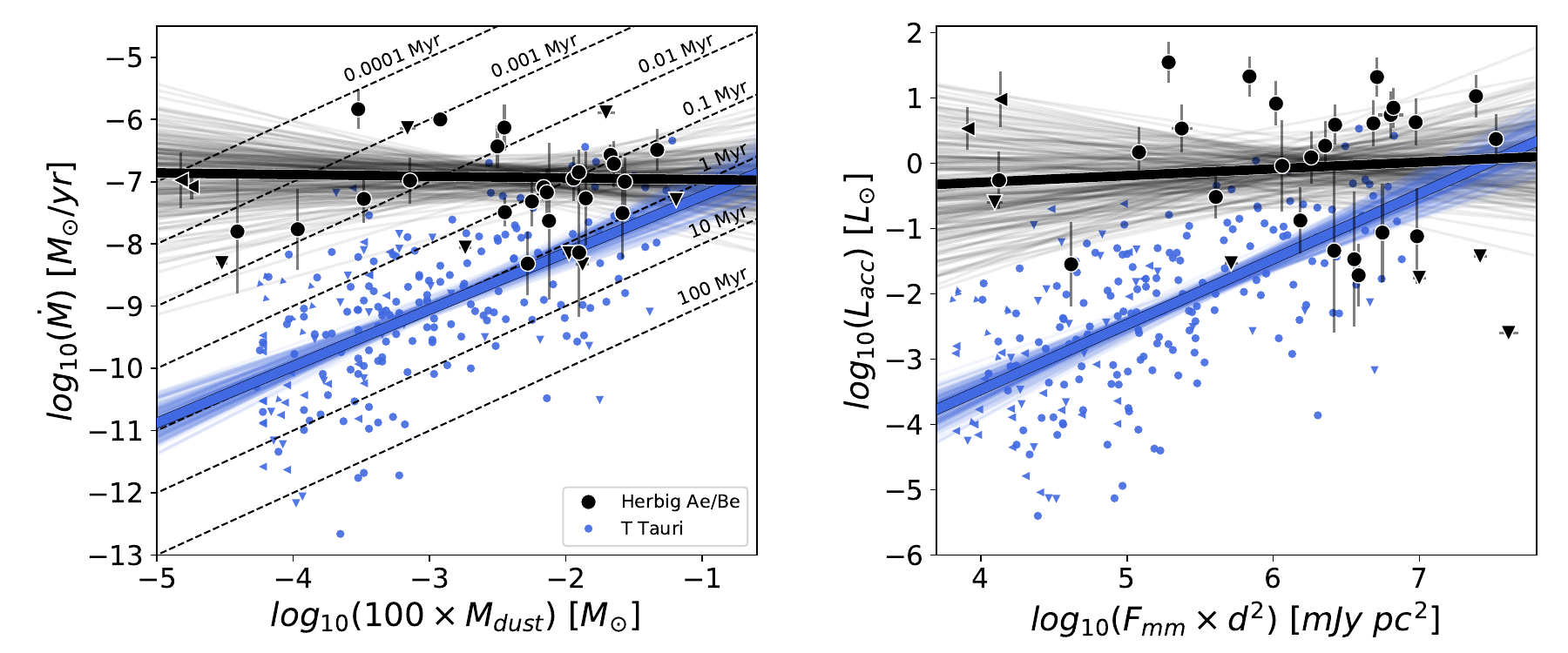}
    \caption{Left: The \mdotmdisk\ relationship for our sample of Herbig Ae/Be stars (black points) and the T Tauri stars from \cite{testi22} (blue points). We have excluded the few sources in the \cite{testi22} sample that have \mstar$>$1.5 \Msun. The black line is the best fit for the Herbig Ae/Be sample and the blue line is the best fit for the T Tauri sample. The thin black lines and thin blue lines are 200 samples of the posterior for the fits to the Herbig Ae/Be and T Tauri points, respectively. Upper limits on the dust mass are shown as leftward facing triangles and upper limits on the accretion rate are shown as downward facing triangles. If both the accretion rate and dust mass measurements are upper limits, the triangle points to the lower left. The dotted gray lines show different disk lifetimes. The Herbig Ae/Be disks are outliers above the already large scatter seen for the lower mass stars. Right: The relationship between the accretion luminosity and the millimeter flux (normalized by the distance).}  
\label{fig: mdot mdisk}
\end{figure*}

\section{Sample, mass accretion rates, and dust masses}\label{sec: samp, mdot, and mdisk}

\subsection{Sample}\label{subsec: sample}

Our sample is compiled from the ALMA-observed sample of \cite{stapper22}, which provides the dust masses used in this work. Their sample consists of the Herbig Ae/Be systems in \cite{vioque18} that are within 450 pc and had available ALMA observations (see \citealt{stapper22} for more details and notes on some excluded objects).

The stellar properties for our sample are listed in Table~\ref{tab: properties} and are largely from \cite{vioque18}. Thirty-one of our 32 sources have high quality Gaia DR2 parallaxes that were used in \cite{vioque18}, which are largely consistent with Gaia (E)DR3 \citep{guzmandiaz21}. One source, HD 53367, was in the low quality sample, and the Gaia DR3 parallax is very different from that of DR2 (parallax of 0.8199$\pm$0.2114 milliarcseconds in DR3 and 7.7682$\pm$0.7854 milliarcseconds in DR2). We keep this source in our sample, using the stellar parameters based on the Gaia DR2 data, but we urge caution in interpreting the results for this source and we do not include it in fits to the \mdotmdisk\ relationship that we present in Section~\ref{sec: results}. The stellar masses range from 1.3 to 16.9 \Msun, but 29 of our 32 sources have stellar masses less than 3 \Msun. Our sample represents a slightly older population, with 27 of our sources having ages greater than 3 Myr. The \cite{meeus01} Group determinations, that are determined from the spectral energy distributions (SEDs) and are thought to reflect the dust disk structure (e.g., \citealt{meeus01,vanboekel05,maaskant13,garufi17,stapper22}), are largely from the SED analysis of \cite{grant22} and \cite{guzmandiaz21}. Our sample is nearly evenly split between Group I (17) and Group II (15) disks.

\subsection{Mass accretion rates}\label{subsec: accretion rates}

The accretion rates for our sample come from the works of \citet{grant22}, \citet{wichittanakom20}, and \citet{garcialopez06}. \cite{grant22} use \brgamma\ observations to derive \Lbrg\ which is then converted to an accretion luminosity using the relationship from \cite{fairlamb17}. Similarly, \cite{wichittanakom20} do the same, except using H$\alpha$ instead of \brgamma. H$\alpha$ and \brgamma\ have a similar spread in the empirical relationship between $L_{\rm{line}}$ and \Lacc\ and are both robust tracers of accretion, even if the line is not generated in the accretion columns (e.g., \citealt{mendigutia15a}). For one object, TY CrA, we use the accretion rate from \cite{garcialopez06}. For this target, the \brgamma\ line is in absorption that is mostly consistent with the photosphere, therefore there is only an upper limit on the accretion rate and we do not include it in the \mdotmdisk\ fits that we discuss in the rest of the paper.

The median accretion rate in our sample is log$_{10}(\dot{M})=-7.09$ (\msunyr), not including upper limits, with a median log error of 0.37. Both sources of the accretion rate measurements rely on the assumption that magnetospheric accretion is the dominant mechanism in these sources. However, the Herbig Ae/Be stellar mass/effective temperature range is thought to be the regime where magnetospheric accretion may break down to boundary layer accretion due to the weak stellar magnetic fields (e.g., \citealt{vink02, donehew-brittain11, mendigutia11b, cauley14,wichittanakom20,grant22}). Based on the findings of \citet{wichittanakom20}, \citet{grant22}, and \citet{vioque22}, the accretion mechanism change may occur at the $\sim$4 \Msun\ boundary and only two stars in our sample above this boundary, HD 53367 and MWC 297.

We have no targets with an accretion rate detection below 10$^{-9}$ \msunyr. One of the criteria needed for Herbig Ae/Be classification is the presence of an accretion tracer, frequently H\,{\textsc{\lowercase{I}}} lines in emission (e.g., \citealt{herbig60,the94}). The use of these lines in identifying Herbig Ae/Be stars is complicated by the fact that these stars have photospheric absorption at those lines and that the depths of the photospheric absorption depends on the stellar effective temperature \citep{joner_hintz15,fairlamb17}. The lower limit on the detectable accretion rate varies with spectral type, the ability to characterize the photosphere, and the measurement method. For example, the lower limit on the measurement of the accretion rate from the veiling of the Balmer jump in the near ultraviolet (NUV) ranges from a few times 10$^{-9}$ \msunyr\ for 2 \Msun\ Herbig stars to about 10$^{-6}$ \msunyr\ for 7 \Msun\ Herbig stars (see Figure 5 in \citealt{sicilia_aguilar16}). If one assumes that the calibration of line luminosity and accretion luminosity inferred from the NUV excess is valid for lower accretion rates, then it is possible to infer lower levels of accretion from spectroscopy of those lines. \cite{fairlamb15} also highlight the changing lower accretion limits based on stellar effective temperature (see their Figure 9). From \cite{sicilia_aguilar16} and \cite{fairlamb15}, an accretion rate of $\sim$10$^{-9}$ \msunyr\ is generally the lower limit for the lowest stellar mass objects in the Herbig Ae/Be classification. The accretion rate values in this work are all above this level, including the outlier objects that are discussed in Section~\ref{sec: results}. We discuss the lack of low accretion rate objects in more detail in Section~\ref{subsec: making sense of the outliers}.

\subsection{Dust masses}\label{subsec: disk masses}
The dust masses in this work were determined in \citet{stapper22} using archival ALMA observations. The spatial resolution in these observations ranges from 0.02\arcsec\ to 1.84\arcsec. Our sample is evenly split between resolved and unresolved disks, although more Group I disks are resolved (11/17) than Group II (5/15). The average spatial resolution for the Group I disks is 0.\arcsec37, while the average is 0.\arcsec88 for the Group II disks. The disk integrated millimeter fluxes were converted to dust masses using a dust temperature that is scaled by the stellar luminosity \citep{andrews13}. The adopted dust opacities, $\kappa_{\nu}$ were determined by a power-law such that $\kappa_{\nu}$=10 cm$^{2}$g$^{-1}$ at 1000 GHz \citep{beckwith90} and scales with an index of 1. In this work we assume that the disk mass is 100 times the dust mass, however, we discuss the implications of this assumption in Section~\ref{subsec: if it is wrong}.

\begin{figure*}
    \centering
    \includegraphics[scale=0.6]{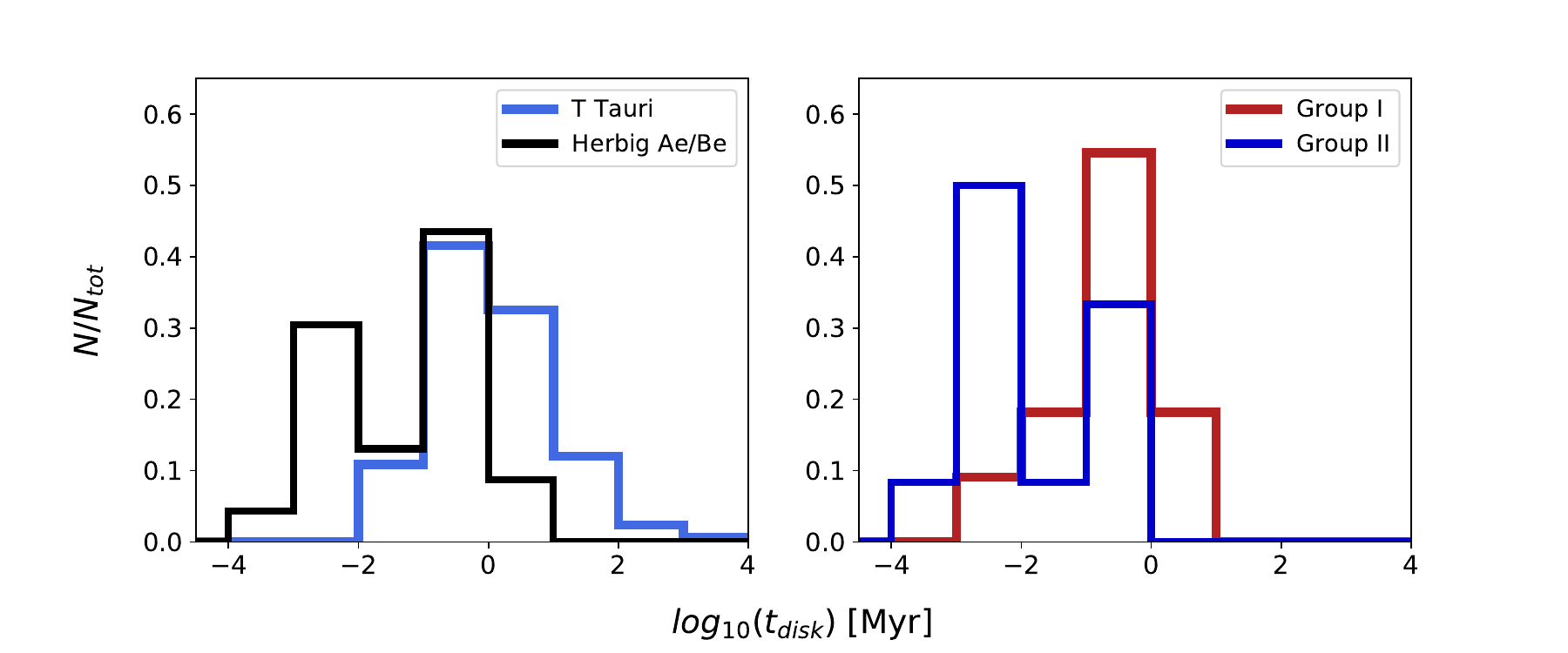}
    \caption{Left: The \tdisk\ distribution for T Tauri disks from \cite{testi22} (blue) and our Herbig Ae/Be sample (black). We have removed any targets from the \cite{testi22} sample that have \mstar$>$1.5 \Msun. Objects with upper limits on \Mdot\ or \Mdisk\ are not included. A two-sample Kolmogorov-Smirnov test returns a p-value of 2.7$\times$10$^{-7}$, indicating that the distributions are drawn from different populations.  Right: The \tdisk\ distribution for Group I disks (red) and Group II disks (blue). A two-sample Kolmogorov-Smirnov test returns a p-value of 0.02, indicating that the distributions may be drawn from different populations. }
    \label{fig: tdisk group}
\end{figure*}

\section{Results}\label{sec: results}

The \mdotmdisk\ relationship for our Herbig Ae/Be sample is presented in Figure~\ref{fig: mdot mdisk}. We fit the Herbig Ae/Be \mdotmdisk\ relationship using the method from \cite{kelly07}\footnote{\url{https://linmix.readthedocs.io}}, taking errors on \Mdot\ and \Mdisk\ and upper limits into account \citep{testi22,fiorellino22}. We find that the \mdotmdisk\ relationship in our Herbig Ae/Be sample is log$_{10}$(\Mdot)=(-0.03 $\pm$ 0.21)log$_{10}$(\Mdisk)+(-6.99 $\pm$ 0.52), a mostly flat relationship over three orders of magnitude in dust mass. This \mdotmdisk\ relationship is drastically different from that of low-mass systems, in particular the large sample compiled and analyzed by \cite{testi22} (Figure~\ref{fig: mdot mdisk}). At the highest disk masses, the Herbig Ae/Be sample largely overlaps with the T Tauri population, although at the higher end of the accretion rate range. However, at the low disk mass end, the Herbig Ae/Be objects lie at and well above the upper end of the T Tauri star accretion rate distribution. The flat relationship that we find for our sample is likely influenced by the fact that Herbig Ae/Be stars require accretion signatures to be classified as such and generally accretion cannot be measured below $\sim$10$^{-9}$ \msunyr\ in intermediate-mass stars. Therefore we lack objects with low accretion rates that may steepen the relationship for intermediate-mass stars in general. While this lower limit is important to keep in mind when interpreting the Herbig Ae/Be \mdotmdisk\ relationship, the flatness of the observed relationship highlights the objects with high accretion rates and low dust masses as clear outliers.

Also shown in Figure~\ref{fig: mdot mdisk} is the relationship between the accretion luminosity and the millimeter flux (normalized by the distance) and the same flat trend is present for the Herbig Ae/Be stars, while the T Tauri stars again show a steeper relationship. The fact that these more ``direct'' quantities show the same relationship indicates that any assumptions going into the determination of the accretion rate and the disk mass (e.g., the dust temperature, magnetospheric accretion being the only source of emission used in determining the accretion rates, etc.) are not the root cause of the flat \mdotmdisk\ relationship for the Herbig Ae/Be sample.

The inferred disk lifetime, $t_{\rm{disk}} = M_{\rm{disk}}/\dot{M}$, is a good measure of how much a given disk deviates from the relationship seen for the T Tauri disks, which cluster around the \tdisk$\sim$1 Myr line. The low disk mass objects in our sample have accretion rates that indicate that the disk will be depleted on much shorter timescales, with 12 of our 32 disks having inferred disk lifetimes of less than 10,000 years (0.01 Myr). We show the distribution of $t_{\rm{disk}}$ in Figure~\ref{fig: tdisk group}, comparing the T Tauri sample of \cite{testi22} to our sample of Herbig Ae/Be sources. A two-sample Kolmogorov-Smirnoff test \citep{scipy} returns a p-value of 2.7$\times$10$^{-7}$, indicating that the T Tauri and Herbig Ae/Be samples are drawn from different populations. Figure~\ref{fig: tdisk group} also shows the Herbig Ae/Be distribution when broken into Group I and Group II sources, showing that the Group II sources are clearly bimodal, while the Group I distribution is unimodal. A two-sample Kolmogorov-Smirnoff test \citep{scipy} returns a p-value of 0.02, indicating that the distributions may be drawn from different populations. We show the \mdotmdisk\ relationship broken up into Group I and Group II objects in Figure~\ref{fig: mdot mdisk group}.

In the Appendix we discuss each of the 12 low disk mass sources that stand out in the \mdotmdisk\ relationship. In particular, we compare our \Mdot\ and \Mdisk\ values to previous values in the literature. We find that our accretion rate values are consistent with those in the literature, subject to differences in accretion determination and variability. In contrast, our disk masses tend to be lower, due to a combination of higher resolution observations, which reduce the amount of contamination from nearby sources/cloud emission, and higher dust temperatures. For instance, this population of high accretion rate, low disk mass objects was not seen in the \mdotmdisk\ analysis of \cite{mendigutia12}, which found that the \mdotmdisk\ relationship for Herbig Ae/Be stars was in line with that of the T Tauri stars. Nine of our objects overlap with their sample and we have compared the accretion rates and dust masses used in each work. Their accretion rates are within an order of magnitude of ours and are evenly split between being higher and lower than our values. The \cite{stapper22} disk mass values are lower than those in \cite{mendigutia12} in 6 sources (one has a higher value in our work, one is an upper limit in \citealt{mendigutia12}, and one has no disk mass determination in \citealt{mendigutia12} due to a lack of millimeter flux). This is due to two differences: 1) the millimeter fluxes from ALMA used by \cite{stapper22} are lower in 6 out of 7 targets, likely due to higher angular resolution observations which suffer less from contamination, and 2) higher dust temperatures used by \cite{stapper22}. The dust temperatures in \cite{stapper22} were determined by scaling by the stellar luminosity while the temperatures in \cite{mendigutia12} were determined from graybody fits to photometry at wavelengths longer than 350 \mic. We note that if we adopt a uniform dust temperature of 20 K, as is commonly done for lower-mass stars, the disk lifetimes increase, but not enough to remove the low-inferred disk lifetimes, with all 12 low disk mass sources having disk lifetimes still less than 0.1 Myr. See the Appendix for further comparison of various disk mass determinations in the literature for these 12 targets.

\begin{figure*}
    \centering
    \includegraphics[scale=0.6]{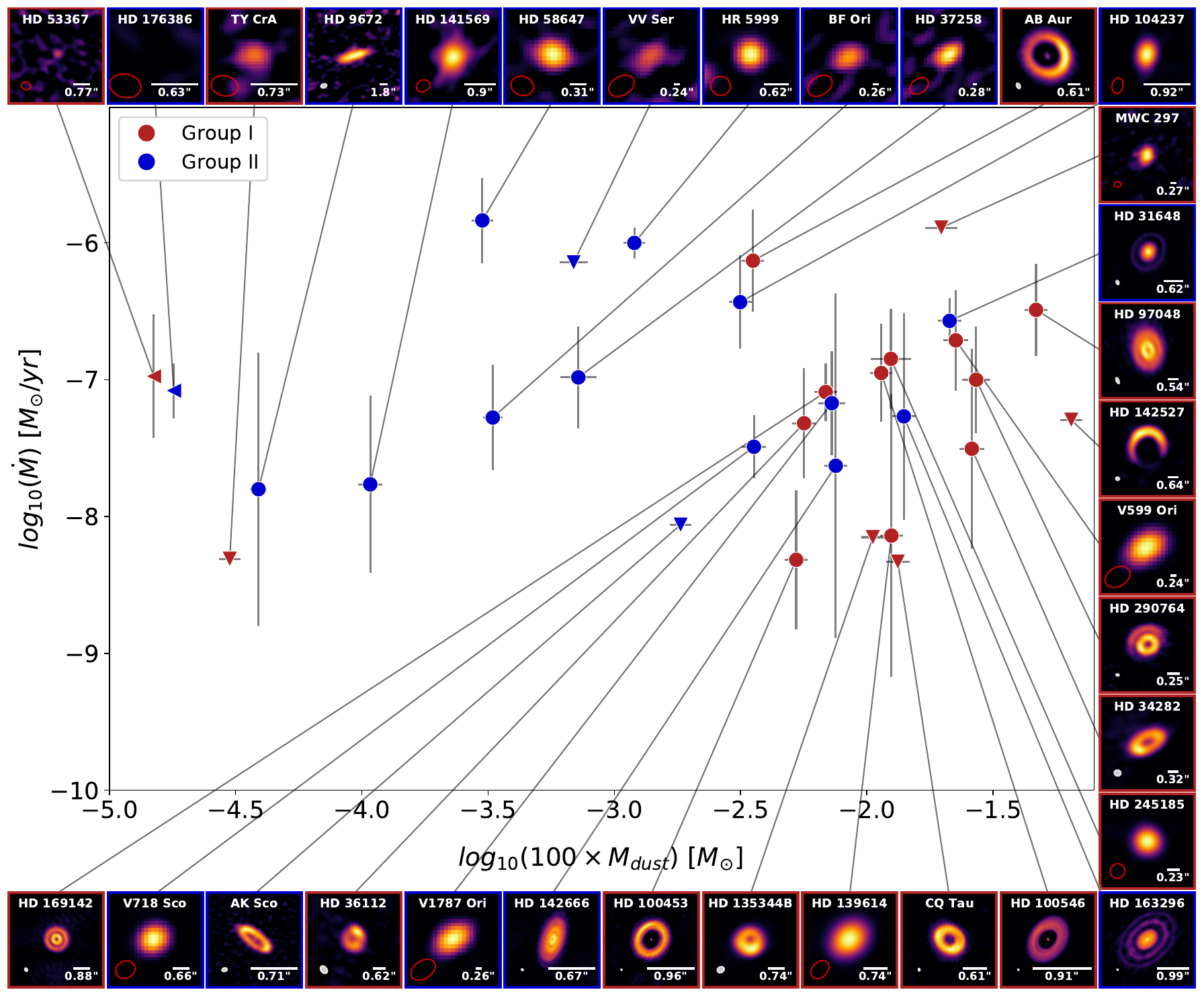}
    \caption{The \mdotmdisk\ relationship for the Herbig Ae/Be stars broken up by group classification (Group I sources in red and Group II sources in blue). Upper limits are the same as in Figure~\ref{fig: mdot mdisk}. 
    ALMA continuum images from \cite{stapper22} are shown for each object with a 100 au scale bar and the beam at the bottom of each image. If the disk is unresolved in the ALMA observations, the beam is shown in red.} 
    \label{fig: mdot mdisk group}
    
\end{figure*}

\section{Discussion}\label{sec: discussion}
In this sample of Herbig Ae/Be objects, we find that the \mdotmdisk\ relationship is relatively flat. While the majority of objects fall along the nominal, steep \mdotmdisk\ relationship of the T Tauri stars, the relationship in our sample is being affected by a subset of objects appearing to have accretion rates inconsistent with their disk masses, such that the disks have a very short inferred lifetime. This outlier sample is likely present due to the biases in Herbig Ae/Be classification, which are limited to objects with accretion rates above $\sim$10$^{-9}$ \msunyr. Here we focus on these short lifetime ``outlier'' objects, first to discuss factors that would move these targets into the nominal \mdotmdisk\ regime, and second how to explain these targets if their disk masses and accretion rates are not under- and overestimated, respectively.

\subsection{Factors that would move the outliers into the general spread}\label{subsec: if it is wrong}
Here, we consider the possibility that either the accretion rates or the dust masses for the low-lifetime objects may be over and under estimated, respectively. 

\begin{itemize}
    
    \item Optical depth: In the scenario where Group II disks are undergoing efficient radial drift, the dust disks will be compact and may be optically thick at millimeter wavelengths which would then lead us to underestimate the dust, and therefore disk, masses \citep{stapper22,liu22}. Modeling efforts, paired with observations at centimeter wavelengths which may be optically thin if the millimeter wavelengths are not, are needed to establish if optically thick emission is the cause of the low disk mass determinations. However, based on the gas masses available for some of these objects (next point), it is unlikely that this is the case for all of these sources.  
    
    \item Gas mass: We are inferring a disk mass based on a gas-to-dust mass ratio of 100. If the true gas-to-dust mass ratio is higher, then our ``low-mass'' disks may be high enough to move the objects to the right enough in the \mdotmdisk\ plane to make the relationship more consistent with what is seen for lower mass stars (e.g., \citealt{sellek20}). Note that the same problem may exist for low-mass stars, even when taking into account the freeze-out of common gas tracers (e.g., \citealt{miotello22}). We find gas mass or gas-to-dust mass ratios available in the literature for 11 of our 32 objects \citep{vandermarel16, boehler17, miley18, yen18, kama20, riviera-marichalar22}. Of these, 5 are upper limits which are above, and therefore consistent with, the disk masses that we use here. There are 4 objects for which the gas mass, or gas-to-dust mass ratios, are below the values inferred from the dust continuum. Finally, two objects have gas masses that are above what we assume here, neither of which changes the disk lifetime substantially. Further discussion of gas masses for the low dust mass, high accretion rate objects is given in the Appendix (Additionally, Stapper et al. in prep will provide a detailed analysis of the gas tracers for this sample). Further careful analysis of gas observations of the disks around Herbig Ae/Be stars is needed to determine the true disk mass, in particular, using gas tracers that are themselves optically thin \citep{booth19}. With these gas masses, we would then be able to determine whether the ``low-mass'' disks are really on the verge of dissipation or whether there is still a large gas reservoir.

    \item Disk winds: There is evidence that disk winds contribute to the \brgamma\ line that is largely used to derive the accretion rates in this work (e.g., \citealt{kraus08b,kurosawa16,hone19,wojtczak22}). If this is the case, then a given accretion rate used here may be artificially inflated. If we instead take accretion rates determined using ultraviolet observations from \cite{donehew-brittain11}, \cite{mendigutia11b}, and \cite{fairlamb15}, which are unaffected by any contribution from a disk wind, the mismatch in slope between T Tauri stars and the Herbig Ae/Be objects is even larger. For example, \cite{mendigutia11b} use the Balmer discontinuity and find an accretion rate of 1.45$\times$10$^{-5}$ \msunyr\ for HD 58647, a factor of 10 higher than the value we use that was determined from \brgamma. On the other hand, \cite{brittain07a} found an accretion rate of 3.5$\times$10$^{-7}$ \msunyr\ using \brgamma. Despite the discrepancies in the accretion rate, none of these values solve the short lifetime implied for this disk which has a dust mass of 1$\pm$0.1 \Mearth. With the \cite{brittain07a} accretion rate the disk lifetime is 860 years, with the \cite{grant22} accretion rate adopted here it is 206 years, and with the Balmer discontinuity accretion rate from \cite{mendigutia11b} it is only 21 years. 

\end{itemize}

To summarize, if all of the outlier Group II disks are extremely optically thick, have gas-to-dust ratios that are much larger than the standard interstellar medium value of 100, or have disk winds that contribute significantly to the accretion tracers used to determine the accretion rate, these objects could really be in the nominal disk lifetime regime. While this needs to be investigated further, literature values of the gas mass and accretion rates determined from ultraviolet observations, which do not suffer from contributions from disk winds, indicate that the trends we are seeing are robust.

\subsection{Making sense of the outliers}\label{subsec: making sense of the outliers}

If the mass accretion rates, dust masses, and gas-to-dust ratios are not wildly off due to the factors discussed above, how might we explain this low-lifetime population of disks? Either these sources are rapidly depleting their disks and we are observing them just as they are about to dissipate, or we are witnessing these sources undergoing variable accretion and happen to be catching them at a point of high accretion that will then decrease before the disk is fully dissipated. We explore these options here.

The low disk lifetime objects are predominantly Group II disks. Our understanding of what these group classifications means has evolved significantly with additional observations and analysis since the classification by \cite{meeus01}. \cite{maaskant13}, \cite{garufi17}, and \cite{stapper22} all find evidence for large cavities in the disks of Group I objects. Additionally, \cite{stapper22} find that the Group I disks have higher dust masses than Group II disks, with Group II disks potentially unable to form giant planets at large radii, resulting in efficient radial drift and compact disks. This agrees with the interpretation of \cite{kama15} and \cite{guzmandiaz23}, who found that refractory elements were depleted in the photospheres of Group I objects relative to Group II disks, suggesting dust trapping in Group I disks by giant planets. The dust mass difference is the source of the difference in the \mdotmdisk\ relationship, as the accretion rates have been found to be consistent between Group I and II systems \citep{mendigutia12,banzatti18,grant22}. 

If we apply the interpretations of Group I and Group II disks as being gapped and potentially hosting giant planets at large radii vs. being unable to form giant planets at large radii and thus having radially compact disks, then the difference in the \mdotmdisk\ relationship becomes more clear (Figure~\ref{fig: mdot mdisk group}). In this scenario Group I disks form giant planets, clearing large gaps in the gas and dust and are surrounded by dust rings at large radii (see the ALMA continuum images in Figure~\ref{fig: mdot mdisk group}). If Group II disks are not able to form giant planets, then they are unable to trap gas or dust in the outer disk, resulting in a rapid inflow of material to the inner disk which maintains a high accretion rate. It is unclear when these systems will then begin to decrease in accretion rate and how rapid that decrease is. This would result in radially compact dust disks for the Group II sources, but higher resolution observations are needed to confirm, as none of the low disk lifetime ($<$0.01 Myr) Group II disks are currently resolved (Figure~\ref{fig: mdot mdisk group}). The Group II disks have an average spatial resolution in the ALMA observations of 0.\arcsec88, compared to 0.\arcsec37 for the Group I disks. Additionally, comparing the gas and dust radii will be crucial for determining if efficient radial drift can explain these systems (e.g., \citealt{trapman19,toci21}).

Other factors that can result in the low disk masses for these objects could come from outer disk depletion from photoevaporation and/or due to multiplicity. If these objects are close to nearby massive stars, the extreme irradiation environments can strip away material, leaving the outer disk depleted (e.g., \citealt{mann14,ansdell17,eisner18,winter18}). Multiplicity has also been shown to impact outer disk evolution, resulting in truncation of the disk \citep{manara19a,panic21, zagaria22}. If any companions are massive stars themselves, then these disks may doubly suffer from truncation and photoevaporation. 

Twenty of the 32 objects in our sample are known binaries, however, the fraction could be higher given the limited surveys that have searched for multiple systems. To identify the binaries in this sample, we use the binary information from \cite{vioque18}, largely collected from \cite{leinert97}, \cite{baines06}, and \cite{wheelwright10}. There are two interesting examples in our sample to study of the effects of multiplicity and photoevaporation: the TY CrA/HD 176386 and HR 5999/HR 6000 systems. TY CrA is in a close triple, if not quadruple, system (e.g., \citealt{vanko13}), and is close to HD 176386, another target in our sample which has a low inferred disk lifetime, and is also a binary. The second example comes from HR 5999, which is itself a binary and is 45\arcsec\ ($\sim$7000 au at a distance of 158 pc) to HR 6000, an early A-type star with no evidence for a disk \citep{stelzer09}. The low disk masses of these systems may be due to photoevaporation and/or truncation from their companions. High spatial resolution observations, in both the gas and dust, paired with photoevaporation and dynamic truncation models (e.g., \citealt{rosotti18}) will help to distinguish the effects of binarity and photoevaporation in these multiple systems.

If these disks are on the verge of dissipation, why do we see them at all? Either these disks are going through an accretion outburst such that despite their low disk masses, we are still able to classify them as Herbig Ae/Be stars, or these high accretion rate, low disk mass objects make up only a small portion of the young, intermediate-mass young stellar object population.

The fact that low \Mdot\ targets are not in our sample is not surprising: Herbig Ae/Be stars are, in part, identified due to the presence of accretion-tracing lines, namely H\,{\textsc{\lowercase{I}}} lines in emission (e.g., \citealt{herbig60,the94}). In practice, only a handful of Herbig Ae/Be stars have rates lower than 1$\times$10$^{-8}$ \msunyr\ (7/267 in the sample of \citealt{vioque22}, 10/102 in the sample of \citealt{grant22}). \cite{mooley13} searched for such objects in the Taurus star forming region. These authors identify three B-type stars and two A-type stars that are probable members. They identify two other stars that are plausible members. Thus, half of the A and B stars in this star-forming region do not show obvious signatures of accretion. \cite{iglesias23} use a volume-limited sample (out to 300 pc) and find that only six out of 134 targets in their sample of young, intermediate-mass stars (1.5 $\leq$ M$_{*}$ $\leq$ 3.5 \Msun) show the accretion signatures needed to designate them as Herbig Ae/Be stars. These results suggest that there is a significant population of A and B stars in our volume (out to 450 pc) that are analogous to the weak lined T Tauri stars. Therefore, the sample of Herbig Ae/Be objects studied in this work may not be representative of the intermediate-mass young stellar object population as a whole, with most of these objects already having dissipated their disks, and thus not meeting the criteria for Herbig Ae/Be objects. Despite this bias, HD 9672 (49 Cet) in our sample is potentially at an intermediate stage, as it has been characterized in different works as a debris disk \citep{zuckermansong12}, albeit one with a large CO gas content (e.g., \citealt{moor19,higuchi20}), and as a Herbig Ae system \citep{vioque18}. Similarly, HD 141569 in our sample, has been considered a ``hybrid'' disk in the transition phase between a protoplanetary disk and a debris disk \citep{augereau04, miley18, difolco20, gravity21,iglesias23}. These targets may represent the bridge between protoplanetary and debris disks. 

The short disk lifetimes inferred in this work have relied on the assumption that the accretion rate is constant in time. However, young stars are known to be variable, with wide-ranging timescales for variability (see the recent review by \citealt{fischer22}). If these low-lifetime targets are undergoing a period of high accretion that will not last, then the disks may not deplete on the short timescales inferred. This has been seen to impact T Tauri stars \citep{claes22}, however this variability may not be enough to explain the spread in the accretion rates measured for T Tauri stars \citep{manara22}. How this variability might be different for higher mass objects, if it is different at all, is unclear (see the discussion on this topic for Herbig Ae/Be objects in \citealt{brittain23}). Characterizing variability in Herbig Ae/Be objects, and putting them into context with young, diskless A and B stars will be crucial for determining whether the low-lifetime population seen here are simply a subset of the larger population that are undergoing periods of strong accretion and thus are included in Herbig Ae/Be samples.

\section{Summary and conclusions}\label{sec: summary}
We analyze a sample of 32 Herbig Ae/Be objects (1.3 to 16.9 \Msun) to determine the relationship between the accretion rate and the dust disk masses. We find the following:

\begin{enumerate}
    \item The mass accretion rate is roughly constant with disk mass, as probed by the dust mass, for Herbig Ae/Be stars (\Mdot$\sim$10$^{-7}$ \msunyr). This is significantly different from the steeper relationship found for T Tauri stars, likely due in part to the biases in classifying stars as Herbig Ae/Bes. While $\sim$two-thirds of the sample follows the $\dot{M}$--$M_{\rm{disk}}$ relationship of the T Tauri stars, one-third has high accretion rates relative to their dust masses. 
    \item T Tauri stars and Herbig Ae/Be systems show very different disk lifetime ($t_{\rm{disk}} = M_{\rm{disk}}/\dot{M}$) distributions, with $\sim$30-40\% of the Herbig Ae/Be sample having disk lifetimes shorter than 0.01 Myr, with this population being dominated by Group II disks (identified by low infrared excesses). 
    \item If the disk masses are underestimated (due to optical depth effects or a higher-than-expected gas-to-dust mass ratio) or the accretion rates are overestimated (due to contributions to the accretion tracers by winds), the outlier objects may actually reside in the nominal \mdotmdisk\ relationship. However, based on values of the disk gas mass measurements and accretion tracers that cannot be contaminated by winds from the literature, it is unlikely this is the cause of all of the low lifetime disks we are observing.
    \item Unless these objects have extreme variability, the outlier disks are on the verge of dissipation. This may be due to efficient radial drift for Group II objects that may not be able to trap material in the outer disk like Group I disks, photoevaporation, and/or truncation of the outer disk due to multiplicity, all of which can result in low disk masses. 
    \item We have no low disk mass, low accretion rate objects in our Herbig Ae/Be sample, highlighting the bias in identifying these objects, which require accretion signatures and infrared excesses to be considered as such. 
    In particular, the inability to measure accretion rates below $\sim$10$^{-9}$ \msunyr\ in these spectral types limits our ability to characterize the \mdotmdisk\ relationship during the last stages of disk evolution in young, intermediate-mass systems. 
    
\end{enumerate}

Further work is needed to characterize the high accretion rate, low dust mass sample. Future high-resolution ALMA observations of these disks are needed to determine whether these disks are compact. Additionally, the sample of Herbig Ae/Be systems with ALMA observations should be expanded, which would allow us to determine whether the low inferred disk lifetime objects constitute only a small fraction of Herbig Ae/Be systems or whether this population is substantial. Finally, to better understand disk evolution around intermediate-mass stars we should also characterize the precursors of Herbig Ae/Be stars, intermediate-mass T Tauri stars, and their descendants, debris disks, to understand how disks move through this plane from formation to dissipation.

\begin{acknowledgements}
We thank the referee for constructive comments that improved the manuscript. We thank Rens Waters, Beno\^{i}t Tabone, and Giovanni Rosotti for useful discussions that contributed to this work. We thank Allegro, the ALMA Regional Center node in the Netherlands, and Aida Ahmadi in particular, for assistance with processing the ALMA data. Astrochemistry in Leiden is supported by the Netherlands Research School for Astronomy (NOVA), by funding from the European Research Council (ERC) under the European Union's Horizon 2020 research and innovation programme (grant agreement No. 101019751 MOLDISK), and by the Dutch Research Council (NWO) grants 648.000.022 and 618.000.001. Support by the Danish National Research Foundation through the Center of Excellence ``InterCat'' (Grant agreement no.: DNRF150) is also acknowledged.

\end{acknowledgements}


\begin{deluxetable*}{cccccccccc}
\tablewidth{0pt}
\caption{Properties of our sample. Accretion rates with $^{a}$ come from \cite{wichittanakom20}, $^{d}$ from \cite{garcialopez06}, and the rest are from \cite{grant22}. Groups marked with $^{b}$ are from \cite{guzmandiaz21}, $^{c}$ from \cite{boersma09}, and the rest are from \cite{grant22}. Dust masses are from \cite{stapper22}. Binary information is from the compilation in \cite{vioque18}. }\label{tab: properties}
\tablehead{\colhead{Source} & \colhead{RA} & \colhead{Dec} & \colhead{$M_{*}$} & \colhead{log$_{10}(L_{*})$} & \colhead{Age} & \colhead{log$_{10}(\dot{M})$} & \colhead{$M_{\rm{dust}}$} & \colhead{Group} & \colhead{Binary}\\
&  & &  \colhead{[$M_{\odot}$]} &  \colhead{[$L_{\odot}$]} & \colhead{[Myr]} & \colhead{[$M_{\odot}/yr$]} & \colhead{[$M_{\oplus}$]} & &  
}
\startdata
AB Aur	 & 	04:55:45.9	 & 	+30:33:04	 & 	2.152	$_{-0.214}^{+0.359}$	 & 	1.61	$_{-0.21}^{+0.19}$	 & 	4.05	$_{-1.49}^{+1.43}$	 & 	-6.13	$\pm$	0.27	$^{a}$	 & 	11.8	$\pm$	1.2	 & 	I		 & 	Yes	\\
AK Sco	 & 	16:54:44.8	 & 	-36:53:19	 & 	1.401	$_{-0.070}^{+0.070}$	 & 	0.62	$_{-0.01}^{+0.03}$	 & 	8.382	$_{-0.42}^{+1.72}$	 & 	$<$-8.06				 & 	6.1	$\pm$	0.6	 & 	II		 & 	Yes	\\
BF Ori	 & 	05:37:13.3	 & 	-06:35:01	 & 	1.807	$_{-0.090}^{+0.090}$	 & 	1.29	$_{-0.05}^{+0.06}$	 & 	6.38	$_{-0.46}^{+0.32}$	 & 	-7.28	$\pm$	0.39		 & 	1.1	$\pm$	0.1	 & 	II		 & 		\\
CQ Tau	 & 	05:35:58.5	 & 	+24:44:54	 & 	1.468	$_{-0.109}^{+0.189}$	 & 	0.87	$_{-0.12}^{+0.18}$	 & 	8.898	$_{-2.52}^{+2.80}$	 & 	$<$-8.33				 & 	44.2	$\pm$	4.8	 & 	I		 & 	Yes	\\
HD 100453	 & 	11:33:05.5	 & 	-54:19:29	 & 	1.251	$_{-0.063}^{+0.063}$	 & 	0.79	$_{-0.00}^{+0.02}$	 & 	6.528	$_{-0.49}^{+0.45}$	 & 	-8.32	$\pm$	0.51		 & 	17.5	$\pm$	1.8	 & 	I		 & 	Yes	\\
HD 100546	 & 	11:33:25.3	 & 	-70:11:41	 & 	2.055	$_{-0.123}^{+0.103}$	 & 	1.37	$_{-0.05}^{+0.07}$	 & 	5.48	$_{-0.77}^{+1.41}$	 & 	-6.95	$\pm$	0.36		 & 	38	$\pm$	3.9	 & 	I		 & 		\\
HD 104237	 & 	12:00:04.9	 & 	-78:11:35	 & 	1.849	$_{-0.092}^{+0.092}$	 & 	1.33	$_{-0.01}^{+0.04}$	 & 	5.48	$_{-0.4}^{+0.27}$	 & 	-6.43	$\pm$	0.34		 & 	10.5	$\pm$	1.1	 & 	II		 & 	Yes	\\
HD 135344B	 & 	15:15:48.4	 & 	-37:09:16	 & 	1.432	$_{-0.072}^{+0.072}$	 & 	0.79	$_{-0.04}^{+0.03}$	 & 	8.927	$_{-0.91}^{+0.45}$	 & 	$<$-8.15				 & 	35.2	$\pm$	3.8	 & 	I		 & 	Yes	\\
HD 139614	 & 	15:40:46.4	 & 	-42:29:54	 & 	1.481	$_{-0.074}^{+0.074}$	 & 	0.77	$_{-0.01}^{+0.03}$	 & 	14.49	$_{-3.60}^{+1.41}$	 & 	-8.14	$\pm$	1.03		 & 	41.7	$\pm$	4.3	 & 	I		 & 		\\
HD 141569	 & 	15:49:57.7	 & 	-03:55:17	 & 	1.860	$_{-0.093}^{+0.093}$	 & 	1.22	$_{-0.03}^{+0.03}$	 & 	8.616	$_{-1.19}^{+11.38}$	 & 	-7.76	$\pm$	0.65		 & 	0.36	$\pm$	0.04	 & 	II		 & 	Yes	\\
HD 142527	 & 	15:56:41.9	 & 	-42:19:24	 & 	1.613	$_{-0.081}^{+0.124}$	 & 	0.96	$_{-0.00}^{+0.03}$	 & 	6.627	$_{-1.55}^{+0.33}$	 & 	$<$-7.29				 & 	214.9	$\pm$	22.1	 & 	I		 & 	Yes	\\
HD 142666	 & 	15:56:40.0	 & 	-22:01:40	 & 	1.493	$_{-0.075}^{+0.075}$	 & 	0.94	$_{-0.05}^{+0.04}$	 & 	9.33	$_{-0.47}^{+0.77}$	 & 	-7.63	$\pm$	1.26		 & 	25.1	$\pm$	2.6	 & 	II		 & 		\\
HD 163296	 & 	17:56:21.3	 & 	-21:57:22	 & 	1.833	$_{-0.092}^{+0.092}$	 & 	1.20	$_{-0.03}^{+0.06}$	 & 	7.598	$_{-1.22}^{+1.05}$	 & 	-7.27	$\pm$	0.75		 & 	46.7	$\pm$	5	 & 	II		 & 		\\
HD 169142	 & 	18:24:29.8	 & 	-29:46:50	 & 	2.000	$_{-0.128}^{+0.131}$	 & 	1.31	$_{-0.22}^{+0.12}$	 & 	8.984	$_{-3.90}^{+11.02}$	 & 	-7.09	$\pm$	0.21	$^{a}$	 & 	22.9	$\pm$	2.4	 & 	I	$^{b}$	 & 		\\
HD 176386	 & 	19:01:38.9	 & 	-36:53:27	 & 	2.299	$_{-0.299}^{+0.143}$	 & 	1.58	$_{-0.22}^{+0.12}$	 & 	4.05	$_{-0.57}^{+15.95}$	 & 	-7.08	$\pm$	0.2	$^{a}$	 & 	$<$0.06			 & 	II	$^{c}$	 & 	Yes	\\
HD 245185	 & 	05:35:09.6	 & 	+10:01:51	 & 	1.923	$_{-0.096}^{+0.177}$	 & 	1.29	$_{-0.10}^{+0.13}$	 & 	7.643	$_{-2.56}^{+12.36}$	 & 	-6.85	$\pm$	0.36		 & 	41.5	$\pm$	7.6	 & 	I		 & 	Yes	\\
HD 290764	 & 	05:38:05.3	 & 	-01:15:22	 & 	1.691	$_{-0.085}^{+0.128}$	 & 	1.18	$_{-0.09}^{+0.09}$	 & 	6.89	$_{-1.41}^{+0.54}$	 & 	-7.0	$\pm$	0.39		 & 	90.3	$\pm$	11.8	 & 	I		 & 		\\
HD 31648	 & 	04:58:46.3	 & 	+29:50:37	 & 	1.779	$_{-0.089}^{+0.131}$	 & 	1.27	$_{-0.05}^{+0.14}$	 & 	6.201	$_{-1.12}^{+0.31}$	 & 	-6.57	$\pm$	0.17	$^{a}$	 & 	70.9	$\pm$	7.7	 & 	II	$^{b}$	 & 		\\
HD 34282	 & 	05:16:00.5	 & 	-09:48:35	 & 	1.450	$_{-0.072}^{+0.072}$	 & 	0.98	$_{-0.04}^{+0.05}$	 & 	6.54	$_{-0.63}^{+2.41}$	 & 	-7.5	$\pm$	0.73		 & 	86.8	$\pm$	9.7	 & 	I		 & 	Yes	\\
HD 36112	 & 	05:30:27.5	 & 	+19:25:57	 & 	1.564	$_{-0.078}^{+0.108}$	 & 	1.04	$_{-0.08}^{+0.12}$	 & 	8.289	$_{-1.40}^{+0.41}$	 & 	-7.32	$\pm$	0.4		 & 	18.8	$\pm$	2	 & 	I		 & 	Yes	\\
HD 37258	 & 	05:36:59.3	 & 	-06:09:16	 & 	1.881	$_{-0.108}^{+0.136}$	 & 	1.24	$_{-0.10}^{+0.12}$	 & 	7.929	$_{-2.45}^{+12.07}$	 & 	-6.98	$\pm$	0.37		 & 	2.4	$\pm$	0.4	 & 	II		 & 	Yes	\\
HD 53367	 & 	07:04:25.5	 & 	-10:27:16	 & 	12	$_{4-}^{+4}$	 & 	3.13	$_{-0.17}^{+0.23}$	 & 	0.08	$_{-0.08}^{+0.08}$	 & 	-6.97	$\pm$	0.45		 & 	$<$0.05		 	 & 	I		 & 	Yes	\\
HD 58647	 & 	07:25:56.1	 & 	-14:10:44	 & 	3.867	$_{-0.193}^{+0.333}$	 & 	2.44	$_{-0.09}^{+0.11}$	 & 	0.8372	$_{-0.18}^{+0.12}$	 & 	-5.84	$\pm$	0.31		 & 	1	$\pm$	0.1	 & 	II		 & 	Yes	\\
HD 9672	 & 	01:34:37.9	 & 	-15:40:35	 & 	1.810	$_{-0.090}^{+0.090}$	 & 	1.17	$_{-0.02}^{+0.09}$	 & 	6.89	$_{-0.51}^{+0.34}$	 & 	-7.8	$\pm$	1.0	$^{a}$	 & 	0.13	$\pm$	0.01	 & 	II	$^{b}$	 & 		\\
HD 97048	 & 	11:08:03.2	 & 	-77:39:17	 & 	2.252	$_{-0.135}^{+0.113}$	 & 	1.54	$_{-0.06}^{+0.07}$	 & 	4.37	$_{-0.32}^{+1.11}$	 & 	-6.49	$\pm$	0.34		 & 	155.9	$\pm$	16	 & 	I		 & 	Yes	\\
HR 5999	 & 	16:08:34.3	 & 	-39:06:19	 & 	2.432	$_{-0.122}^{+0.122}$	 & 	1.72	$_{-0.04}^{+0.05}$	 & 	2.729	$_{-0.35}^{+0.26}$	 & 	-6.0	$\pm$	0.11	$^{a}$	 & 	4	$\pm$	0.4	 & 	II	$^{b}$	 & 	Yes	\\
MWC 297	 & 	18:27:39.5	 & 	-03:49:52	 & 	16.901	$_{-1.215}^{+1.868}$	 & 	4.59	$_{-0.12}^{+0.12}$	 & 	0.02754	$_{-0.006}^{+0.006}$	 & 	$<$-5.89				 & 	65.7	$\pm$	9.6	 & 	I		 & 	Yes	\\
TY CrA	 & 	19:01:40.8	 & 	-36:52:34	 & 	2.063	$_{-0.190}^{+0.223}$	 & 	1.41	$_{-0.23}^{+0.14}$	 & 	6.38	$_{-2.01}^{+13.62}$	 & 	$<$-8.31			$^{d}$	 & 	0.10	$\pm$	0.01	 & 	I	$^{b}$	 & 	Yes	\\
V1787 Ori	 & 	05:38:09.3	 & 	-06:49:17	 & 	1.659	$_{-0.083}^{+0.094}$	 & 	1.15	$_{-0.09}^{+0.11}$	 & 	7.43	$_{-1.05}^{+0.59}$	 & 	-7.17	$\pm$	0.38		 & 	24.2	$\pm$	2.9	 & 	II		 & 		\\
V599 Ori	 & 	05:38:58.6	 & 	-07:16:46	 & 	2.029	$_{-0.101}^{+0.101}$	 & 	1.44	$_{-0.06}^{+0.06}$	 & 	4.289	$_{-0.54}^{+0.42}$	 & 	-6.71	$\pm$	0.37		 & 	75	$\pm$	8.6	 & 	I		 & 		\\
V718 Sco	 & 	16:13:11.6	 & 	-22:29:07	 & 	1.605	$_{-0.080}^{+0.080}$	 & 	0.90	$_{-0.04}^{+0.05}$	 & 	9.804	$_{-0.49}^{+2.80}$	 & 	-7.49	$\pm$	0.23	$^{a}$	 & 	11.9	$\pm$	1.3	 & 	II	$^{b}$	 & 	Yes	\\
VV Ser	 & 	18:28:47.9	 & 	+00:08:40	 & 	2.892	$_{-0.145}^{+0.145}$	 & 	1.95	$_{-0.08}^{+0.10}$	 & 	2.77	$_{-0.21}^{+8.13}$	 & 	$<$-6.14				 & 	2.3	$\pm$	0.3	 & 	II		 & 		\\
\enddata
\end{deluxetable*}

\clearpage

\bibliographystyle{aasjournal}
\bibliography{biblio.bib} 

\begin{appendix}

In this section, we compare the \Mdot\ and \Mdisk\ values that we use in this work to previous values in the literature for all of the objects with an inferred disk lifetime less than 0.01 Myr. In general, we find that high spatial resolution (sub-)millimeter observations are needed to properly determine the disk dust masses, especially for targets with nearby companions which may contaminate low-resolution observations. 

\begin{itemize}

\item AB Aur\\
\cite{guzmandiaz21} find that AB Aur has a disk mass of 0.009$\pm$0.002 \Msun, assuming a gas-to-dust ratio of 100, using a variety of (sub-)millimeter observations. This is within a factor of three of our value of 0.0035 \Msun (\Mdust=11.8 \Mearth). \cite{riviera-marichalar22} find that the gas-to-dust mass ratio varies in the disk of AB Aur, from $\sim$10-40. Therefore, the disk mass is likely to be lower than what we use here, resulting in an even lower disk lifetime than we infer. 

There are several values of the accretion rate for AB Aur in the literature: log$_{10}$(\Mdot)=–6.85 (\msunyr) \citep{garcialopez06}, log$_{10}$(\Mdot)=-7.74 (\msunyr) \citep{donehew-brittain11}, log$_{10}$(\Mdot)=-6.90 (\msunyr) \citep{salyk13}, and log$_{10}$(\Mdot)=-6.13 (\msunyr) \citep{wichittanakom20}. We adopt the value from \cite{wichittanakom20}. 

\item HD 104237\\
\cite{guzmandiaz21} find a disk mass of 0.008$\pm$0.002 \Msun\ for HD 104237, a factor of less than three larger than our value of 0.003 \Msun\ (\Mdust=10.5 \Mearth). The \cite{guzmandiaz21} disk mass is based on 1.27 mm observations from \cite{henning94b} using the 15 m SEST telescope with a resolution of 23\arcsec, which would contain several additional sources in the beam. The ALMA observations are not high enough resolution to resolve the disk, but a companion is observed in the continuum, indicating that we are resolving out some, if not all, sources of additional contamination. \cite{kama20} using HD observations from \textit{Herschel}/PACS observations find a gas-to-dust mass ratio of $\leq$300. 

HD 104237 hosts a binary pair at the center of the circumbinary disk. This has resulted in interesting work on the nature of the inner disk in this system. \cite{garcia13} find that \brgamma\ is variable, with the line equivalent width changing by a factor of 2 depending on the binary interaction, however \cite{garcialopez06} find an accretion rate of log$_{10}$(\Mdot)=–7.45 (\msunyr), one order of magnitude lower than the value we use here. 

\item HD 37258\\
\cite{vanTerwisga22} find a dust mass of 8.9$\pm$0.41 \Mearth\ for HD 37258, relative to the one derived by \cite{stapper22} of 2.4$\pm$0.4 \Mearth. These measurements are from the same ALMA observations (2019.1.01813.S, PI: S. van Terwisga), but the fluxes derived are slightly different and the dust temperatures are different, with \cite{vanTerwisga22} assuming T$_{dust}$=20 K and \cite{stapper22} using T$_{dust}$=51 K, derived from the stellar luminosity. 

\cite{fairlamb15} find an accretion rate of log$_{10}$(\Mdot)=-6.98 (\msunyr), the same value as found by \cite{grant22}.

\item BF Ori\\
\cite{guzmandiaz21} report a disk mass of 0.005$\pm$0.002 \Msun\ for BF Ori using observations from the IRAM 30 m telescope presented in \cite{natta97}. This is 15 times higher than our adopted value from ALMA observations of 3.3$\times$10$^{-4}$ \Msun\ (\Mdust=1.1 \Mearth). This is due to two reasons, the first is that the IRAM observations have a main beam width of 11\arcsec, which likely suffer from contamination compared to the ALMA observations which have a beam of 1.\arcsec49$\times$1.\arcsec03. The ALMA observations are still not high enough resolution to resolve the disk, but are high enough to minimize contamination from nearby objects and cloud contamination. The second contributing factor is the choice of dust temperature, with \cite{guzmandiaz21} using a temperature of 25 K and \cite{stapper22} using a value of 58 K. 

Several works have reported accretion rates for BF Ori: log$_{10}$(\Mdot)=-7.06 (\msunyr) \citep{donehew-brittain11}, log$_{10}$(\Mdot)$<$-8.0 (\msunyr) \citep{mendigutia11b}, log$_{10}$(\Mdot)=-6.65 (\msunyr) \citep{fairlamb15}, and log$_{10}$(\Mdot)=-7.28 (\citealt{grant22}; adopted here). The \cite{donehew-brittain11}, \cite{mendigutia11b}, and \cite{fairlamb15} values are all determined using the Balmer excess, which are 0.22, 0.0, and 0.15 mag, for each of those works, respectively. BF Ori is known to exhibit UX Ori-type behavior (e.g., \citealt{shenavrin12}), with photometric variability in the visual and infrared, therefore we adopt the most recent measurement for the accretion rate, which is within the spread of the previous measurements.

\item HR 5999\\
\cite{guzmandiaz21} find a disk mass of 0.008$\pm$2.19$\times$10$^{-4}$ \Msun, using observations from SCUBA \citep{sandell11} and SMA \citep{meeus12}, in comparison to the disk mass we adopt here of 0.0012 \Msun. The derived fluxes are quite similar with \cite{meeus12} deriving a 1.3 mm flux of 34.3$\pm$0.9 mJy and \cite{stapper22} deriving a flux of 26.5 mJy. A companion is seen in the ALMA continuum observations which is well-resolved from HR 5999. \cite{yen18} used $^{13}$CO and C$^{18}$O ALMA observations, paired with models from \cite{miotello16}, to determine a gas mass of 6$_{-3.2}^{+7.2}\times$10$^{-5}$ \Msun\ for HR 5999, which is 20 times lower than our estimate from the dust mass. This gas mass value, when taken with an accretion rate of log$_{10}$(\Mdot)=-6.0 (\msunyr), results in an inferred disk lifetime of only 60 years.

\cite{wichittanakom20} re-derived the accretion rate for HR 5999 from the observations of \cite{fairlamb15,fairlamb17} with updated stellar parameters, finding an accretion rate of log$_{10}$(\Mdot)=-6.0 (\msunyr), compared to the accretion rate by \citet{fairlamb15} of log$_{10}$(\Mdot)=-6.25 (\msunyr). We adopt the accretion rate from \cite{wichittanakom20}. 

\item VV Ser\\
\cite{guzmandiaz21} determine a disk mass of 9.54$\times$10$^{-4}\pm$2.730$\times$10$^{-4}$ \Msun\ for VV Ser using observations from the Plateau de Bure Interferometer \citep{alonso-albi08,boissier11} with a beam of 1.\arcsec7$\times$0.\arcsec8 at 1.3 mm. This matches well with our derived disk mass of 6.9$\times$10$^{-4}$ \Msun\ (2.3 \Mearth\ in dust mass). \cite{pontoppidan07a} found that the mass of the small dust grains is only $\sim$0.03 \Mearth. 

\cite{mendigutia11b} find a Balmer jump (0.54 mag) that is inconsistent with magnetospheric accretion models, however \cite{donehew-brittain11} find a Balmer jump of 0.16 mag, which is within the range of magnetospheric models run by \cite{mendigutia11b}. \cite{garcialopez16} find that several H\,{\textsc{\lowercase{I}}} lines, including \brgamma, are variable, with \brgamma\ likely to originate in a disk wind. The models used in that work assume an accretion rate of 3.3$\times$10$^{-7}$ \msunyr. With these discrepant measurements, it is unclear whether the accretion mechanism is variable, if magnetospheric accretion is taking place in this object, and if so, how much of the \brgamma\ line is generated from magnetospheric accretion. However, while the accretion mechanism may be unclear, the accretion rate is likely to be high, given the high Balmer jump observed in both \cite{mendigutia11b} and \cite{donehew-brittain11}. We note that \cite{donehew-brittain11} find an accretion rate of log$_{10}$(\Mdot)=-7.49 (\msunyr), but with a pre-Gaia distance and stellar properties. We adopt an upper limit to the accretion rate of log$_{10}$(\Mdot)$<$-6.14 (\msunyr) from \cite{grant22} and do not include it in the \mdotmdisk\ fits.   

\item HD 58647\\
Few (sub-)millimeter observations of HD 58647 are available in the literature. We consider the ALMA observations (from Program 2018.1.00814.S), with an RMS of 0.14 mJy beam$^{-1}$ and a beam of 0.\arcsec47$\times$0.\arcsec39, and the dust mass determination of 1$\pm$0.1 \Mearth\ from \cite{stapper22} to be robust. 

In comparison to the lack of (sub-)millimeter observations, HD 58647 has several U-band and near-infrared observations. \cite{mendigutia11b} use the Balmer discontinuity and find an accretion rate of log$_{10}$(\Mdot)=–4.84 (\msunyr) for HD 58647. Using \brgamma\ observations, \cite{brittain07a} find an accretion rate of log$_{10}$(\Mdot)=-6.45 (\msunyr), \cite{ilee14} find an accretion rate of log$_{10}$(\Mdot)=-6.32 (\msunyr), and \cite{grant22} find an accretion rate of log$_{10}$(\Mdot)=-5.84 (\msunyr). We adopt the latter in this work.

\item HD 141569\\
\cite{miley18} find a gas mass of 6$\times$10$^{-4}$ \Msun\ for HD 141569 using ALMA $^{13}$CO (2-1) observations, a factor of 6 above our inferred disk mass of 1$\times$10$^{-4}$ \Msun. This higher disk mass is still low enough that the inferred disk lifetime is only 0.03 Myr. \cite{guzmandiaz21} find a disk mass of 1.46$\times$10$^{-4}$ \Msun\ for HD 141569 from millimeter observations, in good agreement with the value that we adopt from \cite{stapper22}.

For HD 141569 several accretion rates have been determined in the literature: \cite{garcialopez06} find an accretion rate of log$_{10}$(\Mdot)=-8.37 (\msunyr), \cite{mendigutia12} find a value of log$_{10}$(\Mdot)=-6.89 (\msunyr), and \cite{fairlamb15} find a value of log$_{10}$(\Mdot)=-7.65 (\msunyr). \cite{grant22} and \cite{wichittanakom20} update the value from \cite{fairlamb15} to log$_{10}$(\Mdot)=-7.76 (\msunyr) and log$_{10}$(\Mdot)=-7.23 (\msunyr), respectively. We adopt the value from \cite{grant22}.

\item HD 9672\\
\cite{moor19} find a CO gas mass of 1.11$\times$10$^{-2}$ \Mearth\ (3.33$\times$10$^{-8}$ \Msun) for HD 9672/49 Cet, however, CO may not be a good tracer of the total disk mass, in particular depending on the gas origin (e.g., \citealt{moor19}). Using the dust continuum and assuming a gas-to-dust mass ration of 100, \cite{guzmandiaz21} find a disk mass of 2.92$\times$10$^{-4}$ \Msun\ from infrared photometry from \textit{Herschel}/PACS, a factor of seven higher than our value of 3.9$\times$10$^{-5}$ \Msun. The ALMA observations analyzed in \cite{stapper22} should provide a more accurate estimate of the dust mass due to the longer wavelength observations. 

The accretion rate of log$_{10}$(\Mdot)=-7.80 (\msunyr) for HD 9672 comes from \cite{wichittanakom20}, derived from the Fiber-fed Extended Range Optical Spectrograph (FEROS, \citealt{kaufer99}) spectra from ESO Program 082.A-9011(A).

\item TY CrA \\
\cite{cazzoletti19} find a dust mass for TY CrA of 0.66 \Mearth\ (disk mass of 2$\times$10$^{-4}$ \Msun, assuming a gas-to-dust mass ratio of 100), compared to that derived by \cite{stapper22} of 0.1 \Mearth\ (disk mass of 3$\times$10$^{-5}$ \Msun, assuming a gas-to-dust mass ratio of 100), from the same dataset, largely due to the difference in dust temperature assumed. \cite{guzmandiaz21} find a disk mass of less than 0.017 \Msun, from the upper limits on the millimeter flux from \cite{henning94b} and \cite{pezzuto97}. 

The only accretion rate in the literature that we found is that of \cite{garcialopez06}, who found an accretion rate of log$_{10}$(\Mdot)$<$–8.31 (\msunyr), based on the depth of the \brgamma\ line, which indicated little or no accretion taking place in this object.

\item HD 176386\\
HD 176386 is undetected in the ALMA observation, with the continuum only present at the 1.6$\sigma$ level \citep{stapper22}. That observation had an RMS of 0.20 mJy/beam, meaning that the upper limit for the flux of 0.32 mJy, corresponding to a dust mass of 0.06 \Mearth\ (1.8$\times$10$^{-5}$ \Msun\ in total disk mass assuming a gas-to-dust ratio of 100). \citet{guzmandiaz21} used sub-millimeter observations from SCUBA \citep{difrancesco08} to determine a disk mass of 0.121$\pm$0.01 \Msun. This nearly 4 orders of magnitude difference is due to contamination in the SCUBA maps, which have a 14\arcsec\ full width at half maximum in the 850 \mic\ map used. HD 176386B is a binary companion to HD 176386 with a separation of 3.\arcsec7 \citep{wilking97}, and would have contributed to the flux observed in the low resolution SCUBA observations. The high resolution ALMA observations, with a beam of 0.\arcsec43$\times$0.\arcsec32, is able to spatially distinguish the sources. 

HD 176386 has several accretion rates in the literature. \cite{garcialopez06} find an accretion rate of log$_{10}$(\Mdot)=-8.11 (\msunyr), \cite{wichittanakom20} find a value of log$_{10}$(\Mdot)=-7.08 (\msunyr), and \cite{guzmandiaz21} find a value of log$_{10}$(\Mdot)=-6.49 (\msunyr). \cite{pogodin12}, whose X-Shooter data is the source of the H$\alpha$ equivalent width used in \cite{wichittanakom20}, conclude that magnetospheric accretion cannot be applied to this object, given that the disk may be dispersed. Given that we have stringent upper limits on the disk mass for this object, we agree that the disk may be dispersed and the accretion rate should be considered with caution. However, the \cite{pogodin12} absorption H$\alpha$ profile for HD 176386 does show shallowing relative to a photospheric model, which may be due to accretion. We take the adapted value from \cite{wichittanakom20}, however this value should be used with caution.

\item HD 53367\\
The ALMA observations analyzed in \cite{stapper22} (from Program 2018.1.00814.S) show that HD 53367 is only present in the continuum at the 2.4$\sigma$ level with an RMS of 0.14 mJy beam$^{-1}$. We are not aware of other (sub-)millimeter observations of this target. 

\cite{donehew-brittain11} find an accretion rate of log$_{10}$(\Mdot)$<$-7.92 (\msunyr) from a Balmer discontinuity of $<$0.09 mag. \cite{fairlamb15} find a similar Balmer discontinuity of 0.10 mag and they are unable to determine an accretion rate for this source given the very high stellar effective temperature of 29500$\pm$1000 K. The \Mdot\ value that we adopt here of log$_{10}$(\Mdot)=-6.97 (\msunyr) is from \cite{grant22}, based on strong \brgamma\ line emission. However, given the high stellar mass and effective temperature of this object, this accretion rate should be viewed with caution. Given the low dust mass of this object, if any accretion is taking place, the disk would likely be depleted very quickly.

\end{itemize}

\end{appendix}

\end{document}